\RequirePackage{etoolbox}\csdef{input@path}{{style/}{graphics/}}
\documentclass[aoas,MSNbibl,nameyear,seceqn,dvips]{arximspdf}
\usepackage{dcolumn}
\usepackage{graphicx}

\doi{10.1214/14-AOAS769} % kopijuoti is laisko
\volume{8}
\issue{4}
\pubyear{2014}
\firstpage{2268}
\lastpage{2291}
\docsubty{FLA}

\makeatletter
\newcolumntype{d}[1]{D{.}{.}{#1}}

\renewcommand{\mid}{|}
\newcommand{\eqref}[1]{(\ref{#1})}
\newcommand{\logit}{\operatorname{logit}}
\newcommand{\dirichletd}{\operatorname{Dirichlet}}
\newcommand{\gammad}{\operatorname{Gamma}}
\newcommand{\discreted}{\operatorname{Discrete}}
\newcommand{\bernoullid}{\operatorname{Bern}}
\newcommand{\indep}{\mathrm{indep}}
\newcommand{\DOB}{\mathrm{DOB}}
\newcommand{\GAM}{\mathrm{GAM}}
\newcommand{\AUG}{\mathrm{AUG}}
\newcommand{\Age}{\mathrm{Age}}
\newcommand{\iid}{\mathrm{i.i.d.}}
\newcommand{\bY}{\mathbf{Y}}
\newcommand{\bx}{\mathbf{x}}
\newcommand{\bZ}{\mathbf{Z}}
\newcommand{\bg}{\mathbf{g}}
\newcommand{\balpha}{{\bolds{\alpha}}}
\newcommand{\bbeta}{{\bolds{\beta}}}
\newcommand{\bxi}{{\bolds{\xi}}}
\makeatother

\begin{document}
\begin{frontmatter}

\title{Longitudinal Mixed Membership trajectory models for disability
survey data\thanksref{T1}}
\runtitle{Mixed Membership trajectory models}

\begin{aug}
\author{\fnms{Daniel}~\snm{Manrique-Vallier}\corref{}\ead[label=e1]{dmanriqu@indiana.edu}}
\runauthor{D. Manrique-Vallier}
\affiliation{Indiana University}
\address{Department of Statistics\\
Indiana University\\
Bloomington, Indiana 47408\\
USA\\
\printead{e1}}
\end{aug}
\thankstext{T1}{Supported in part by NIH Grant R01 AG023141-01 to
Carnegie Mellon University and NSF Grant SES-11-31897 to Duke University.}

% HISTORY:
\received{\smonth{10} \syear{2012}}
\revised{\smonth{4} \syear{2014}}

% ABSTRACT
%
\begin{abstract}
We develop methods for analyzing discrete multivariate longitudinal
data and apply them to functional disability data on the U.S. elderly
population from the National Long Term Care Survey (NLTCS), 1982--2004.
Our models build on a Mixed Membership framework, in which individuals
are allowed multiple membership on a set of extreme profiles
characterized by time-dependent trajectories of progression into
disability. We also develop an extension that allows us to incorporate
birth-cohort effects, in order to assess inter-generational changes.
Applying these methods, we find that most individuals follow
trajectories that imply a late onset of disability, and that younger
cohorts tend to develop disabilities at a later stage in life compared
to their elders.
\end{abstract}

% KEYWORDS
\end{frontmatter}

\setcounter{footnote}{1}
%s1 #&#
\section{Introduction}
This paper introduces new models and estimation procedures to analyze
discrete multivariate longitudinal data on functional disability,
motivated by the analysis of data from the National Long Term Care
Survey (NLTCS). The NLTCS is a longitudinal panel survey instrument
aimed at assessing chronic disability among the elderly (65$+$)
population in the United States. It enables researchers to answer
important questions related to the aging process and disability
prevalence in the U.S.: how many elder Americans will live with
disabilities? What is the of duration of disability episodes? What is
the age of onset of disability? Is the nature of disability changing
for younger generations?
[\citet{connor2006trn}]. Answers to these questions are of
importance in public policy design due to, among other reasons, the
increased public and private expenditure for disabled people in
contrast with their able peers [\citet{mantonlambwu2007}].

Many of the relevant public policy questions for which the NLTCS can
potentially provide answers are related to changes over time: changes
during the life of an individual (``how is this individual likely to
age?'') or comparing people across different generations (``are people
from later generations acquiring disabilities differently than people
born 20 years before?''). Thus, to answer these questions we need to
consider the longitudinal dimension of these data. In addition, as not
everyone could be expected to age the exact same way, it is safe to
assume that elderly American people constitute a heterogeneous
population. Models for longitudinal disability data need to be capable
of accounting for such heterogeneity.

Although the longitudinal nature of the NLTCS data is frequently
invoked [e.g., \citet
{cordermanton1991,mantonstallard1997,mantonetal2006}], efforts to
analyze the data using true longitudinal methods have been few and far
between. Most researchers have instead analyzed the NLTCS as a series
of uncorrelated cross-sectional samples [see, e.g., \citet
{mantonstallard1997,mantonetal2006,mantonlambwu2007}].
Recent attempts to deal with the longitudinal nature of the NLTCS have
been undertaken by \citet{stallard2005}, \citet
{connor2006} and \citet{whiteerosheva2013}.

The new models and methods that we propose in this paper, which we call
\textit{Trajectory Grade of Membership} models (TGoM), seek to capture
both the longitudinal nature of the individual NLTCS data and the
inherent individual heterogeneity of the aging process. These models
handle individual heterogeneity using the concept of Mixed Membership
[\citet{erosheva2004,eroshevafienberg2005}]. Mixed Membership
models describe a small number of ideal types of individuals (or
extreme profiles) and let each individual partially belong to each pure
type, to a different degree. At the same time, TGoM models focus on the
longitudinal nature of the process by defining the extreme profiles as
typical progressions over time. We also introduce an extension to this
model aimed at capturing differences across generational cohorts. We do
this by allowing individuals' Mixed Membership to depend on their dates
of birth.

The remainder of this article is organized as follows. In the next
section we present a brief introduction and description of the National
Long Term Survey. Next, in Section~\ref{secmodel}, we describe the
basic TGoM model and its extension to handle generational cohorts.
Estimation algorithms based on MCMC sampling are introduced in
Section~\ref{secestimation} and fully described in Appendices~\ref{secmcmcbasic} and~\ref{secmcmcgrouped}. In Section~\ref
{secapplication} we apply the TGoM models to the NLTCS. Finally, in
Section~\ref{secdiscussion}, we conclude with a discussion on the
insights provided by the models, their limitations and possible extensions.

%s2 #&#
\section{The National Long Term Care Survey}
\label{secintroNLTCS}
The National Long Term Care Survey (NLTCS) is a longitudinal panel
survey designed specifically to assess the state and progression of
chronic disability among the United States population aged 65 years or
more [\citet{cordermanton1991}]. It consists of six waves,
conducted in 1982, 1984, 1989, 1994, 1999 and 2004. In very rough
terms, each wave consists of interviews to approximately 20,000 people,
from which around 15,000 are previously interviewed individuals. Each
wave includes a fresh new sample of around 5000 individuals. These
refreshment samples serve the double purpose of replacing those who
have died since the previous wave and of keeping each wave
representative of the current state of the population over 65
[\citet{Clark1998}]. A total of around 49,000 people have been
screened in the survey between 1982 and 2004.

The NLTCS assesses functional disability by evaluating subjects'
ability to perform two sets of activities. The first one, called \textit{Activities of Daily Living} (ADL), comprises basic self-care
activities, such as bathing, eating and dressing. The second, \textit{Instrumental Activities of Daily Living} (IADL), involves activities
necessary for independent living within a community, like preparing
meals or maintaining finances. The NTLCS determines the functional
status in these activities through answers to a series of triggering
questions, which are then summarized as binary response items that
indicate the presence or absence of impairments.

The design of the NLTCS is such that the survey data can be used as
several cross-sectional samples, considering each wave as a different
sample from the target population at that time, and also as a
longitudinal sample, following individuals across different measurement waves.

The NLTCS first screens each sampled individual using a special,\break
``screener,'' questionnaire aimed at quickly detecting if he or she is
chronically disabled. The operational definition of ``chronically
disabled'' in the context of the NLTCS requires that the individual
presents an impairment in some ADL or IADL lasting or expected to last
at least 90 days. If screened-out, the individual's status is
registered and they are re-screened in subsequent waves, to assess if
the disability status has changed. If the individual is screened-in, he
or she is then interviewed using a detailed questionnaire. There are
different detailed questionnaires for institutionalized and individuals
living in the community. After receiving a detailed questionnaire for
the first time, the subject is then eligible to receive detailed
questionnaires in all subsequent waves of the survey until death
[\citet{Clark1998}].

In what follows, we have used a subset of the NLTCS consisting of all
six binary answers to questions about the individual's ability to
perform ADLs (EAT:~Eating; DRS:~Dressing; TLT:~Toileting; BED:~Getting in
and out of bed; MOB:~Inside mobility; BTH:~Bathing), from all six waves
of the NLTCS. We obtained ages and dates of birth from linked Medicare
data from the Centers for Medicare and Medicaid services (CMS). We
provide further details about our data preprocessing in Section~\ref{secapplication}.

%s3 #&#
\section{Mixed Membership trajectory models}
\label{secmodel}
The goal of this analysis is to characterize typical progressions in
acquisition of disabilities over time while taking into consideration
and characterizing the heterogeneity of the population. For this we
combine two main ideas.

The first idea is clustering based on trajectories. This is the idea
behind Latent Trajectory models [LTMs; \citet{nagin1999}].
Broadly speaking, LTMs are mixture models of the form
%
%e3.1 #&#
\begin{equation}
p(\mathbf{y}\mid \bx) = \sum_{k=1}^K
\pi_k f_k(\mathbf{y}\mid \bx),
\end{equation}
where $\mathbf{y}$ is a vector containing $T$ longitudinal
measurements of a response variable of interest and $\bx$ is a vector
that contains the corresponding $T$ values of a time-dependent
covariate. The joint densities corresponding to each mixture component,
$f_k(\cdot)$, are in turn modeled using parametric \textit{trajectory
functions}. Trajectory functions (or simply \textit{trajectories})
describe typical progressions over time, usually modeling the
dependence of the outcome variables as a function of age.
For a given population, LTMs provide estimates of both the trajectories
and the individuals' distribution over them. Therefore, LTMs perform
data-driven clustering based on evolution over time [see \citet
{nagin1999} for details]. \citet{connor2006} adapted this
technique for the analysis of multivariate discrete data and applied it
to the NLTCS. The trajectory curves represented the probability of
presenting a disability as a function of age. This tool provides a
simple and easy mechanism to interpret typical ways of aging, with a
degree of heterogeneity handling. However, it assumes that individuals
within a class are perfectly homogeneous. It thus attributes all the
potential within-class variability to random fluctuations. In Connor's
formulation, this assumption essentially says that, within a class,
every single individual responds to the exact same underlying aging
process. It thus disregards the fact that classes are ideal
constructions to which possibly no actual individuals belong
[\citet{kreutermuth2008}].

The second idea, Mixed Membership, provides a powerful and conceptually
attractive way of relaxing the within-class homogeneity assumption.
Similarly to traditional clustering techniques, like the Latent Class
model [\citet{goodman1974}] or LTMs, Mixed Membership models
still assume the existence of a small number of classes, called \textit{ideal types} or
\textit{extreme profiles}. However, instead of forcing
every individual into one and only one class, they allow them to belong
simultaneously to more than one, in different degree. The Grade of
Membership model [GoM; \citet
{woodburyclivegarson1978,mantonstallardwoodbury1991}] is an
example of a Mixed Membership model that has been successfully applied
to the cross-sectional analysis of the NTLCS [see, e.g., \citet
{mantonstallard1997,mantonetal2006,mantonlambwu2007,eroshevafienbergjoutard2007}].
\citet{eroshevafienbergjoutard2007} developed a full Bayesian
version of the GoM model and applied it to a pooled across-waves
version of the NLTCS.

The approach we present here combines LTMs with Mixed Membership. It
seeks to produce a soft clustering based on trajectories. Similar to
LTM, it assumes that for a given population we can identify a few ways
of progressing over time, which we consider ideal extreme cases. At the
same time it assumes that individuals in the population do not exactly
correspond to these typical profiles, but instead behave somewhere in
between them, in quantifiable ways.
Note that this approach is conceptually different from previous
cross-sectional applications of the GoM model to the study of
disability. In those applications extreme profiles represented ideal
types of disability, whereas in TGoMs they represent ideal types of
people. In the same way, it also differs from other previously proposed
time-dependent Mixed Membership models, which specify time-evolving
individual membership [e.g., \citet
{stallard2005,xing2010state}]. In TGoMs the membership is an immutable
characteristic of the individual.

%s3.1 #&#
\subsection{Basic TGoM model}\label{secbasicconstruction}
We consider a sample composed of $N$ individuals. Following Mixed
Membership ideas, we assume the existence of a number, $K$, of
reference types of individuals called \textit{extreme profiles}. These
extreme profiles represent idealized individuals. This means that it
might be the case that no real individual corresponds exactly to any of
them. Instead, we assume that each individual $i = 1,\ldots,N$ has an
associated \textit{membership vector}, $\bg_i = (g_{i1},\ldots,g_{iK})$,
whose $k$th component, $g_{ik}$, represents their \textit{degree of
membership} into the $k$th extreme profile. We constrain membership
vectors so that their components are positive numbers that sum to 1,
that is, they lie on a $(K-1)$-dimensional unit simplex, $\Delta_{K-1}$.
In this way, we identify ideal individuals of the $k$th type as those
whose membership vectors' components are zeros on each component
distinct from $k$, and $g_{ik}=1$. For instance, we say that an
individual with membership vector $\bg_i = (0,0,1,0)$ belongs
exclusively to the extreme profile $k = 3$. Similarly, we can represent
more complex membership structures. For example, $\bg_i = (0.1, 0.2,
0.4, 0.3)$ indicates that individual $i$ has 10\% membership in the
first extreme profile, $20\%$ in the second and so on.

We are interested in modeling the progression of disability as time
passes. We start by modeling ideal individuals. Let individual $i$
provisionally be a full member of extreme profile $k$, that is, $g_{ik}
=1$. Let $y_{ij}(\tau)$ be $1$ if the individual does experience
difficulties performing ADL $j$ at age $\tau$, and $0$ otherwise. We
model the evolution of the probability of a positive response to
question $j$, $y_{ij}(\tau)$, as a function of age, $\lambda
_{jk}(\tau)$, so that
%
%e3.2 #&#
\begin{equation}
\lambda_{jk}(\tau) = \Pr \bigl(y_{ij}(\tau) = 1\mid
g_{ik}=1, \bbeta _{jk}, \tau \bigr).
\end{equation}
Here $\bbeta_{jk}$ is a generic vector of parameters that indexes
$\lambda_{jk}(\cdot)$ within a parametric family, for example, the
parameters of a linear logistic curve. We call the functions $\lambda
_{jk}(\cdot)$ \textit{extreme trajectories}.

Now moving to actual individuals, we specify the corresponding
trajectory of a generic, nonideal individual $i$, with membership
vector $\bg_i = (g_{i1},\ldots,g_{iK})$, as the convex combination
\begin{eqnarray*}
\label{eqbasicabstractindivtraj} \lambda^{(i)}_j(\tau) &=& \Pr
\bigl(y_{ij}(\tau) = 1\mid \bg_i, \bbeta _j,
\tau \bigr)
= \sum_{k=1}^K g_{ik}
\lambda_{jk}(\tau),
\end{eqnarray*}
where $\bbeta_{j}=(\bbeta_{j1},\ldots,\bbeta_{jK})$.\vspace*{1pt}

Although $\tau$ is a continuously-varying quantity, we only have
measurements at each of the $t=1,\ldots,T=6$ occasions, corresponding to
the waves of the survey. Thus, we define $y_{ijt} = y_{ij}(\Age_{it})$,
where $\Age_{it}$ is the age of individual $i$ at measurement time $t =
1,\ldots,T$.
We group these numbers into individual vectors $\mathbf{Age}_i =
(\Age_{i1},\ldots,\Age_{iT})$.
Then we have that
\begin{eqnarray*}
p(y_{ijt}\mid \bg_i, \bbeta_j,
\mathbf{Age}_i) &=& \bernoullid \bigl(y_{ijt}\mid
\lambda^{(i)}_j(\Age_{it}) \bigr)
\\
&=& \sum_{k=1}^Kg_{ik}
\bernoullid \bigl(y_{ijt}\mid\lambda _{jk}(\Age_{it})
\bigr),
\end{eqnarray*}
where $\bernoullid(y\mid p)=p^y(1-p)^{1-y}$, for $y \in\{0,1\}$ and $0<p<1$.

\begin{figure}

\includegraphics{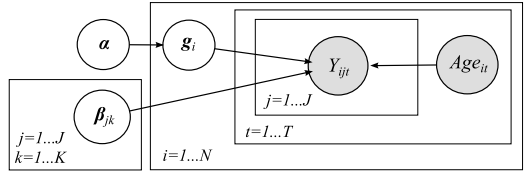}

\caption{Graphical probabilistic representation of the basic TGoM
model. Observed variable $\Age_{it}$ is the age of individual $i$ at
survey wave $t$. Gray nodes represent observed quantities; white nodes
represent parameters to estimate.}
\label{figtgomgraphical}
\end{figure}

Next, we assume that, for a single individual, the $J$ responses at
each of the $T$ measurement times are conditionally independent of one
another, given their membership vector, $\bg_{i}$, and covariate
vector $\mathbf{Age}_i$. Under this assumption we effectively use the
membership vector and the covariates to decouple the dependence
structure present in the components of the response. Then we have
%
%e3.3 #&#
\begin{eqnarray}
\label{eqbasicabstractindivlikelihood} p(\mathbf{Y}_i\mid \bg_i, \bbeta,
\mathbf{Age}_i)&=& \prod_{j=1}^{J}
\prod_{t=1}^T \sum
_{k=1}^K g_{ik} \bernoullid
\bigl(y_{ijt}\mid \lambda _{jk}(\Age_{it})\bigr),
\end{eqnarray}
where $\mathbf{Y}_i= (y_{ijt})_{J=1,\ldots, J, t=1,\ldots, T}$ and $\bbeta=
(\bbeta_1,\ldots,\bbeta_J)$.
By assuming that each individual has been randomly sampled from the
population, we finally get the joint model of $\mathbf{Y} = (\mathbf
{Y}_i)$, conditional on $\bg= (\bg_i)$ and $\mathbf{Age} = (\mathbf
{\mathbf{Age}}_i)$,
%
%e3.4 #&#
\begin{eqnarray}
p ( \mathbf{Y} \mid \bg, \bbeta, \mathbf{Age} ) &=& \prod
_{i=1}^{N} \prod_{j=1}^J
\prod_{t=1}^{T} \sum
_{k=1}^K g_{ik} \bernoullid
\bigl(y_{ijt}\mid \lambda _{jk}(\Age_{it})\bigr).
\end{eqnarray}
We assume that membership vectors are i.i.d. samples from a common
distribution~$G_{\balpha}$, with support on the simplex $\Delta
_{K-1}$. This yields the unconditional (on $\bg$) model for the sample
$\bY$,
%
%e3.5 #&#
\begin{eqnarray}
\qquad p (\bY\mid \bbeta, \mathbf{Age} ) &=& \prod
_{i=1}^{N} \int_{\Delta_{K-1}} \prod
_{j=1}^J \prod
_{t=1}^T \sum_{k=1}^K
\omega_k \bernoullid\bigl(y_{ijt}\mid \lambda_{jk}(\Age_i)
\bigr) G_\balpha(d\bolds \omega),
\end{eqnarray}
where $\bolds\omega= (\omega_1,\ldots,\omega_K) \in\Delta_{K-1}$.
Figure~\ref{figtgomgraphical} shows a graphical representation of
the structure of this model.
%
%f1 #&#

As \citet{eroshevafienbergjoutard2007} discuss for the Grade
of Membership model, the model in \eqref
{eqbasicabstractindivlikelihood} admits the augmented data representation,
%
%e3.6 #&#
\begin{eqnarray}\label{eqbasicaugmentedabstract}
&& f^{\AUG} (\bY_i, \bZ_i \mid
\mathbf{Age}_i, \bbeta, \bg_i )
\nonumber\\[-8pt]\\[-8pt]\nonumber
&&\qquad = \prod
_{j=1}^J \prod_{t=1}^{T}
\prod_{k=1}^K \bigl[g_{ik}
\bernoullid\bigl(y_{ijt}\mid \lambda_{jk}(\Age_{it})
\bigr) \bigr]^{I(z_{ijt}=k)},
\end{eqnarray}
where $\bZ_i = (z_{ijt})_{j=1,\ldots, J, t=1,\ldots, T}$ with $z_{ijt} \in\{
1,2,\ldots,K\}$.
Following \citet{eroshevafienbergjoutard2007},
it is easy to show that the expression in \eqref
{eqbasicabstractindivlikelihood} is equivalent to
%
%e3.7 #&#
\begin{equation}
p (\bY_i \mid \mathbf{Age}_i, \bbeta,
\bg_i ) = \sum_{\mathbf{z}\in\mathcal{Z}}f^{\AUG} (
\bY_i, \mathbf {z}\mid \mathbf{Age}_i, \bbeta,
\bg_i ),
\end{equation}
where $\mathcal{Z}= \{(z_{jt})_{J\times T}\dvtx  z_{jt} \in\{
1,\ldots,K\}  \}$. We then see that the model in \eqref
{eqbasicabstractindivlikelihood} can be thought of as a
marginalized version of the model in \eqref
{eqbasicaugmentedabstract}. This equivalence shows that the TGoM
model conforms to the general mixed-membership structure described in
\citet{erosheva2004}. It also
makes it possible to construct algorithms for posterior inference of
the TGoM using the augmented model [\citet{tanner1996}].

%s3.2 #&#
\subsection{Detailed specification}
\label{secbasiccontinuoustrajectories}
The extreme trajectories functions, $\lambda_{jk}(\cdot)$, encode
several assumptions about the dynamics of the underlying process over
time. Thus, their specific functional form must be
application-specific. For this application to the NLTCS, following
\citet{connor2006}, we use a linear logit specification%
%e3.8 #&#
\begin{eqnarray}
\label{eqlogisticcurve} \logit \bigl[ \lambda_{jk}(\tau) \bigr] &=&
\beta_{0jk} + \beta _{1jk} \tau.
\end{eqnarray}
Here $\bbeta_{jk} = (\beta_{0jk}, \beta_{1jk})$. This specification
expresses the intuitively sound notion that the underlying probability
of disability is a monotonic function of age. It also has the advantage
of being relatively simple, with just $2 \times J$ parameters per
extreme profile. In the supplementary material [\citet{supp}] we present an analysis using an
alternative specification and include a discussion about the
appropriateness of \ref{eqlogisticcurve}.

Similar to \citet{erosheva2002} and \citet
{airoldibleifienberg2008}, we take the common distribution of the
$N$ membership vectors $\bg_i$, $G_\balpha$, as
%
%e3.9 #&#
\begin{equation}
\bg_{i} \mid \balpha\mathop{\sim}\limits^{\iid} \dirichletd(\balpha),
\end{equation}
where $\balpha= (\alpha_1, \alpha_2,\ldots,\alpha_K)$ with $\alpha
_k>0$ for all $k=1,\ldots,K$.

The Dirichlet distribution has some good properties in this setting.
First, it is conjugate to the multinomial distribution. This simplifies
computations using Gibbs samplers. Second, adopting the
reparametrization $\balpha= (\alpha_0 \cdot\xi_1,\ldots,\alpha_0
\cdot\xi_K)$ with $\alpha_0>0$, $\xi_k >0$ and $\sum_k \xi_k =
1$, we can interpret the vector $\bxi=(\xi_1,\ldots,\xi_K)$ as the
average proportion of responses generated by the $k$th extreme profile
and $\alpha_0$ as a parameter governing the spread of the
distribution: as $\alpha_0$ approaches $0$, the samples from
$G_\balpha$ are more and more concentrated on the vertices of the
simplex $\Delta_{K-1}$; and as $\alpha_0$ increases they are more
concentrated near its mean, $\bxi$.

As \citet{eroshevafienbergjoutard2007} and \citet
{airoldifienbergetal2007} discuss, a~priori setting parameter
$\balpha$ in the Dirichlet distribution is too strong an assumption to
do realistic modeling. Estimates can be highly sensitive to this prior
specification. For this reason, we prefer to estimate these parameters
directly from the data, specifying hyperpriors and computing posterior
distributions. We specify hyperpriors for $\alpha_0$ and $\bxi$
similar to \citet{erosheva2002} and \citet
{eroshevafienbergjoutard2007}: $\alpha_0 \sim\gammad(a_\alpha,b_\alpha)$ and $\bxi\sim\dirichletd(\mathbf{1}_K)$. This
specification takes advantage of the interpretation of the parameters
$\alpha_0$ and $\xi$, considering them as independent entities and
modeling them separately. For the same reason we also assume that
$p(\alpha_0, \bxi) = p(\alpha_0)p(\bxi)$.

We specify the priors for the parameters that define the extreme
trajectories, $\bolds{\beta}_{jk} = (\beta_{0jk}, \beta
_{jk})$, as two independent normal distributions, $\beta_{0jk} \mathop
{\sim}\limits^{\iid}\break  N(\mu_0, \sigma_0^2)$ and $\beta_{1jk} \mathop
{\sim}\limits^{\iid} N(\mu_1, \sigma_1^2)$, for all $j =1,\ldots,J$ and
$k=1,\ldots,K$. These priors can be set to be noninformative, by a
priori specifying high variances. We also assume that $\bolds
\beta_{jk}$ are a priori independent of $\balpha$.

%s3.3 #&#
\subsection{Representing generational changes}\label{secgenerational}
The basic TGoM model from Section~\ref{secbasicconstruction} takes
advantage of the longitudinal nature of the NLTCS by following
individuals as they age. It, however, attributes all variation over
time, including changes in prevalence of disability patterns, to the
individual progression of aging. Thus, it attributes all changes in
prevalence of disability between different epochs to the aggregation of
individuals that are at distinct points of their life trajectories.

To answer questions about changes in the ways of aging across different
generations---for example, ``are younger generations acquiring
disabilities differently than older ones?''---we need to take into
account the birth cohort of individuals. We do so by modeling the
dependence between cohorts and the membership scores, keeping the
extreme trajectories the same for the whole population. This
arrangement allows us to read differences in the ways of aging as
differences in the underlying distribution of membership, conditional
on birth cohort. We interpret these differences using the common frame
of reference provided by the extreme trajectories.

A direct way of enabling inter-generational comparisons under this
framework is to keep the individual-level structure proposed for the
basic TGoM model, but replace the common distribution of membership
vectors with a family indexed by a function of the date of birth (DOB)
covariate:
\begin{eqnarray*}
p (y_{ijt} \mid \bg_i, \Age_{it}, \bbeta ) &=&
\sum_{k=1}^K g_{ik} \bernoullid
\bigl(y_{ijt}\mid \lambda _{jk}(\Age_{it}) \bigr),
\\
\bg_{i}\mid \DOB_i &\mathop{\sim}\limits^{\indep} &
G_{\balpha(\DOB_i)}.
\end{eqnarray*}
For our application we keep the Dirichlet specification, but replace
its parameter $\balpha$ with a function of DOB, so that $G_{\balpha
(\DOB)} = \dirichletd(\balpha_{(\DOB)})$.
We note that under this specification the membership vectors, $\bg_i$,
are now dependent on a covariate.

\begin{figure}

\includegraphics{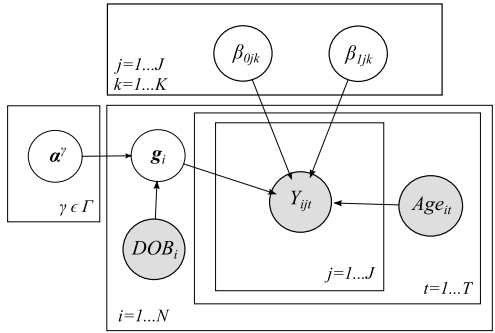}

\caption{Probabilistic graphical representation of the extended TGoM
model with cohort effects. Gray nodes represent observed quantities;
white nodes represent parameters to estimate.}
\label{figtgomcohortgraphical}
\end{figure}

A simple, yet reasonably flexible, way of specifying $\balpha(\DOB)$ is
by defining a number of cohorts and making it constant within each of
them. Let $\Gamma= \{ \gamma_1, \gamma_2,\ldots,\gamma_C\}$ be a
finite partition (contiguous nonoverlapping intervals) of the range of
possible dates of birth. Define $\balpha(\DOB) =  (\alpha_1(\DOB),
\alpha_2(\DOB),\break \ldots, \alpha_K(\DOB) )$ by
%
%
%e3.10 #&#
\begin{equation}
\label{eqgroupedstepspecification} \alpha_k(\DOB) = \prod_{\gamma\in\Gamma}
\bigl( \alpha^\gamma_k \bigr)^{I(\DOB \in\gamma)},
\end{equation}
where $\alpha^\gamma_k>0$. Then, we extend the TGoM model to handle
cohort information by replacing the population level distribution of
membership vectors, $p(\bg_i\mid \balpha)$, with its conditional version,
$p (\bg_i \mid\balpha(\DOB_i) )$. Figure~\ref{figtgomcohortgraphical} shows a graphical representation of this
expanded model.
%
%f2 #&#

We specify the same hyperprior distribution that we used for the basic
TGoM for all the newly introduced parameters. To this end, define
$\alpha^\gamma_0 = \sum_{k=1}^K \alpha^\gamma_k$ and $\xi^\gamma
_k = \alpha^\gamma_k/\alpha^\gamma_0$, and take
$
\alpha^\gamma_0 \mathop{\sim}\limits^{\iid} \gammad(\tau, \eta)$
and $\bxi^\gamma= (\xi^\gamma_1,\ldots,\xi^\gamma_K) \mathop{\sim
}\limits^{\iid} \dirichletd(\mathbf{1}_K)$.
%s4 #&#
\section{Estimation}\label{secestimation}
We developed MCMC algorithms based on Gibbs sampling to obtain samples
from the posterior distribution of parameters for both the basic and
the generational model. These algorithms rely on the augmented data
representation in \eqref{eqbasicaugmentedabstract}. We present the
full description
in Appendices~\ref{secmcmcbasic} and~\ref{secmcmcgrouped}.

%s5 #&#
\section{Application to the NLTCS}\label{secapplication}
We have selected an extract from the NLTCS data that includes data from
all six waves. These data include all the individuals that received the
screener in at least one of the first five waves of the survey (1982,
1984, 1989, 1994 or 1999). We excluded individuals who entered the
sample for the first time in 2004 because of lack of information about
their dates of birth and death. Similarly, we excluded all the
individuals that were institutionalized in 1982 because the NLTCS did
not register their ADL statuses that year.
The resulting sample size was $N={}$38,428 subjects.
For each individual at each wave we focused on six ADLs: Eating
($j=1$), Dressing ($j=2$), Toileting ($j=3$), Getting In or Out of Bed
($j=4$), Inside Mobility ($j=5$) and Bathing ($j=6$). We determined the
age of each individual in years by computing the difference between the
interview and birth dates, and assuming 365 days for all years.
For computing and prior specification purposes, we recentered ages at
80 years. However, for clarity we report any estimates or descriptive
statistics related to age without the offset.

%
%t1 #&#
\begin{table}%[tbp]
\tabcolsep=0pt
\caption{Cohort definition and distribution by wave}\label{tabcohortdefinition}
\begin{tabular*}{\tablewidth}{@{\extracolsep{\fill}}@{}lcd{4.0}d{4.0}d{4.0}d{4.0}d{4.0}d{4.0}@{}}
\hline
&&\multicolumn{6}{c@{}}{\textbf{Wave}}\\[-6pt]
&&\multicolumn{6}{c@{}}{\hrulefill}\\
& & \multicolumn{1}{c}{\textbf{1982}} & \multicolumn{1}{c}{\textbf{1984}} &
\multicolumn{1}{c}{\textbf{1989}} & \multicolumn{1}{c}{\textbf{1994}}&\multicolumn{1}{c}{\textbf{1999}}  &
\multicolumn{1}{c}{\textbf{2004$^{\bolds{*}}$}}
\\
\textbf{Cohort} & \multicolumn{1}{c}{\textbf{DOB}} &\multicolumn{1}{c}{$\bolds{(t=1)}$} &
\multicolumn{1}{c}{$\bolds{(t=2)}$} &
\multicolumn{1}{c}{$\bolds{(t=3)}$} &
\multicolumn{1}{c}{$\bolds{(t=4)}$} &
\multicolumn{1}{c}{$\bolds{(t=5)}$} &
\multicolumn{1}{c@{}}{$\bolds{(t=6)}$}
\\
\hline
 1 &\phantom{0000}--1906& 6329 & 6025 & 1347 & 1397 &  617 &  70 \\
 2 &1906--1914& 7631 & 7082 & 3452 & 3335 & 1753 &  575 \\
 3 &1914--1919& 3696 & 7839 & 2627 & 5102 & 3679 & 2010 \\
 4 &1919--1926&  1 &  463 & 2410 & 4581 & 4724 & 3505 \\
 5 &1926--\phantom{0000}&  0 &  0 &  0 & 2478 & 6403 & 4251 \\
\hline
\end{tabular*}
\tabnotetext{($*$)}{Only individuals present in 1999.}
\end{table}

We defined five cohorts or generational groups, partitioning the ranges
of possible dates of birth according to the intervals defined in the
second column of Table~\ref{tabcohortdefinition}. We selected these
intervals so that they group approximately the same number of
individuals. A salient feature of this arrangement is that individuals
from the youngest cohort (cohort 5---born after 1926) have measurements
only in the last three waves due to age eligibility, as its oldest
members turned 65 after 1991. Also note that neither the oldest (cohort
1---born before 1906) nor the youngest (cohort~5---born after 1926)
cohorts span the whole range of relevant dates of birth in the NLTCS.
In fact, the oldest individual in cohort 5 could be at most 78 years
old in 2004, while the youngest individual from cohort 1 could not be
younger than 76 years old in 1982.

%s5.1 #&#
\subsection{Basic GoM trajectory model}
\label{secresultsbasic}
We fitted the basic model described in Section~\ref
{secbasiccontinuoustrajectories} to the NLTCS data using the MCMC
algorithm from Appendix~\ref{secmcmcbasic}, for $K=2,3,4$ and 5
extreme profiles.

We set the prior distribution for the proportions parameter of the
membership vector, $\bxi$, as a uniform distribution over $\Delta
_{K-1}$, or $\dirichletd(\mathbf{1}_K)$. We specified the prior
distribution for the corresponding concentration parameter, $\alpha
_0$, as $\gammad(1,5)$, in shape/inverse scale parametrization. This
last specification expresses a slight preference for small values of
$\alpha_0$, although not very pronounced. This choice is more a
modeling decision than an expression of prior knowledge: small values
of $\alpha_0$ in the Dirichlet parametrization have the effect of
concentrating the probability of individual membership vectors around
the vertices of the unit simplex. This has the effect of producing
individual membership vectors where one single profile is predominant,
but where the other profiles still exert some effect. This behavior is
a desirable characteristic from an interpretative standpoint that
allows us to discuss ``predominant'' profiles, while still having a
significant degree of flexibility in the handling of heterogeneity due
to the influence of the other profiles. For the parameters governing
the extreme trajectories, we selected diffuse independent normal priors
with $\mu=0$ and variance $\sigma^2 = 100$.

In all cases, the MCMC chains converged rapidly, reaching stationary
distributions after approximately 15,000 iterations. Still, run times
were long due to the chains' slow mixing. In all cases, we ran 120,000
iterations, discarded the first 20,000 and subsampled them, keeping
20\% of the remaining. Similar to other latent variable models, the
TGoM is invariant to permutation of its extreme profile labels. Thus,
we inspected the trace plots for signs of label switching [\citet
{jasraholmes2005labelswitch}]. No switching was found. Although
label switching is a potential problem, in this application the modal
regions of the posterior distributions seem to be well separated due to
the abundance of data.

%s5.1.1 #&#
\subsubsection{Basic model results}\label{secbasiccontinuousresults}
The basic TGoM model includes parameters that represent two distinct
structural features: typical ways of aging, given by the extreme
profiles (parameters $\bolds{\beta}$), and the way individuals
distribute with respect to these extreme profiles (parameters $\bxi$
and $\alpha_0$). Extreme profile parameters can be difficult to
interpret directly. Thus, we instead consider the quantities given by
the transformation
%
%e5.1 #&#
\begin{equation}
\Age_{q, jk} = -\frac{1}{\beta_{1jk}} \biggl[\beta_{0jk} + \log
\biggl(\frac{1-q}{q} \biggr) \biggr] + 80,
\end{equation}
for $q=0.1$, $q=0.5$ and $q=0.9$. These parameters express the age at
which an ideal individual of the extreme profile $k$ reaches a
probability $q$ of being unable to perform ADL $j$. The 80 year offset
is required because we have recentered the age data, subtracting 80. We
also relabel extreme profiles according to the decreasing sequence of
posterior estimates of $\xi_k$. This is necessary because of the
TGoM's invariance to permutations of the extreme profile labels. This
way, the expression ``first extreme profile'' ($k=1$) will always refer
to the extreme profile with the highest relative importance in the
population (the one to which most individuals are the closest; see
Section~\ref{secresultsbasic}) and ``the last'' ($k=K$) to the one
with the lowest.

%
%f3 #&#
\begin{figure}[b]

\includegraphics{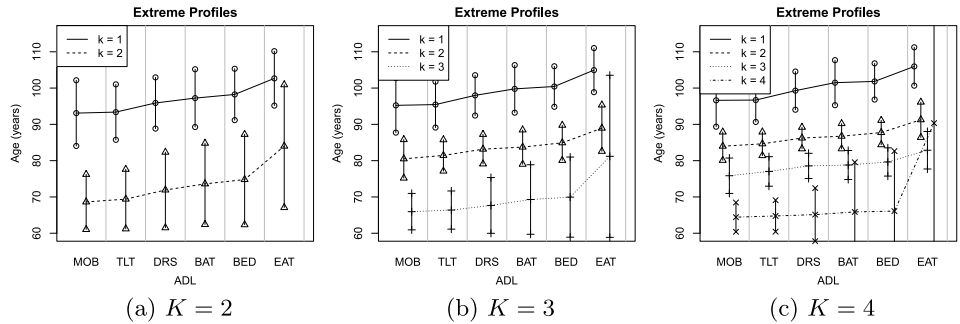}

\caption{Posterior estimates of extreme profiles for models with
$K=2,3,4$. Vertical segments represent the age range at which ideal
individuals' probabilities of disability go up from 0.1 to 0.9, for
each ADL ([$\Age_{0.1,jk}, \Age_{0.9,jk}$]). For visualization purposes,
ADLs are sorted according to $\Age_{0.5,jk}$ posterior estimates.}\label{figbasicprofiles}
\end{figure}

%
%t2 #&#
\begin{table}
\tabcolsep=0pt
\caption{Posterior estimates of population-level parameters for basic
model with $K=2,3,4,5$ extreme profiles. Numbers between parenthesis
are posterior standard deviations}\label{tabbasicalphaestimates}
\begin{tabular*}{\tablewidth}{@{\extracolsep{\fill}}@{}lcccccc@{}}
\hline
& $\bolds{\alpha_0}$ & $\bolds{\xi_1}$ & $\bolds{\xi_2}$ & $\bolds{\xi_3}$ & $\bolds{\xi_4}$ & $\bolds{\xi_5}$\\
\hline
$K=2$ & 0.328 & 0.824 & 0.176 & -- & -- & -- \\
& (0.007) & (0.002) & (0.002) \\
$K=3$ & 0.261 & 0.645 & 0.251 & 0.104 & -- & -- \\
& (0.006) & (0.004) & (0.004) & (0.002)\\
$K=4$ & 0.237 & 0.540 & 0.259 & 0.124 & 0.078 & -- \\
& (0.006) & (0.005) & (0.004) & (0.003) & (0.002)\\
$K=5$ & 0.235 &0.496 & 0.244 & 0.128 & 0.074 & 0.058 \\
& (0.005) &(0.007) & (0.006) & (0.003) & (0.002) & (0.001) \\
\hline
\end{tabular*}
\end{table}

Figure~\ref{figbasicprofiles} and Table~\ref
{tabbasicalphaestimates} present summaries of the posterior
distribution of the extreme profile and Mixed Membership parameters,
respectively. Plots in Figure~\ref{figbasicprofiles} are based on
posterior means of the quantities $\Age_{q,jk}$, for models with
$K=2,3,4$. For each extreme profile, vertical line segments represent
the age interval at which the probability of being unable to perform
each ADL increases from 10\% to 90\%, that is, $[\Age_{0.1,jk},
\Age_{0.9,jk}]$. To aid visualization, we sorted the ADLs according to
the $\Age_{0.5,jk}$ estimates. Note that\vspace*{1pt} this procedure resulted in the
exact same sequence of ADL in every case. Table~\ref
{tabbasicalphaestimates} shows the posterior summaries of the Mixed
Membership distribution parameters, $\alpha_0$ and $\bxi$.

Estimates of the parameter $\alpha_0$, in Table~\ref
{tabbasicalphaestimates}, are relatively small for all models. This
was expected since the prior distribution of $\alpha_0$, $\gammad
(1,5)$, was already expressing strong a priori preference for
small values of $\alpha_0$. However, as we can note from their very
small posterior dispersion relative to the prior dispersion, these
estimates are strongly data driven. This is not surprising, considering
the large amount of data available to perform the estimations.

For all models, the extreme profile with the highest relative
importance in the population, $k=1$, represents a pattern of healthy
aging, with a very late onset of disability. Extreme trajectories in
this profile show that for any ADL, ideal individuals in this class
have a very small probability of experiencing disability until
approximately age 90. The remaining extreme profiles show patterns with
progressively earlier onsets of disability, as we consider the extreme
profiles in sequence. This is a feature worth noting: all models point
to an inverse relationship between the relative importance of a profile
in the population and its implied age of onset of disability. That is,
most people's aging trajectories are closer to a profile that describes
a late onset of disability.

We note that the sequence of ADLs obtained from sorting them according
to their implied age of onset of disability (represented by parameter
$\Age_{0.5,jk}$) is the same for all extreme profiles of all models.
Closer inspection reveals that the pattern of acquisition of
disabilities directly inferred from the data closely follows what we
can interpret as a sequence of activities decreasingly sorted in terms
of difficulty: inside mobility, toileting, dressing, bathing, getting
in and out of bed, and eating.

Another salient feature of these results is that for $k=1, 2, 3$ and
$4$, the inferred slope parameters of the extreme trajectories ($\beta
_{1jk}$) are all positive, even though the prior specification allows
for negative values. This result supports the intuition that the
probability of experiencing a disability in any ADL can only increase
as one ages. It also makes it possible to construct the graphical
summaries in Figure~\ref{figbasicprofiles}. The only exception to
this regularity is in profile $k=5$, in the model with $K=5$ extreme
profiles. In this profile trajectories exhibit a counterintuitive
decreasing progression. We note that the relative importance of this
profile in the population is small, with $\hat\xi_{5} \approx0.058$
(compared with $\hat\xi_1 \approx0.496$ for the most important
profile). From a modeling perspective, an obvious way of avoiding this
type of aberrant behavior is to make it an impossibility a
priori, restricting the support of the slope parameters to positive
values. We have implemented such a model. However, while the rest of
the parameters remained almost the same, the slope parameter of most
trajectories in this profile were zero or very close to zero. These
outcomes---together with the results obtained using a different
trajectory specification, in the supplementary material [\citet{supp}]---suggest that this profile
captures a small residual variability, which is not correctly modeled
by the main extreme trajectories. Accounting for this effect is an area
for future improvements.

\begin{figure}

\includegraphics{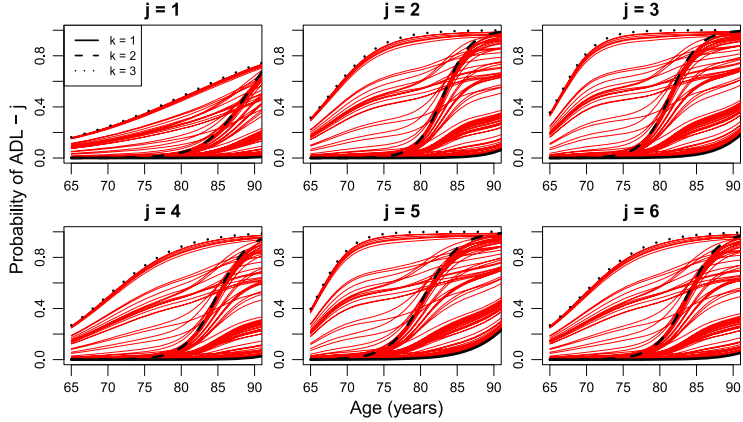}

\caption{Individual-level mixture of trajectories for model with $K=3$
extreme profiles for each ADL. Extreme trajectories are represented
with thick lines and a random sample of 100 individual posterior
trajectory curves are plotted using thin lines.}
\label{figbasicmusclek3}
\end{figure}

\begin{figure}[b]

\includegraphics{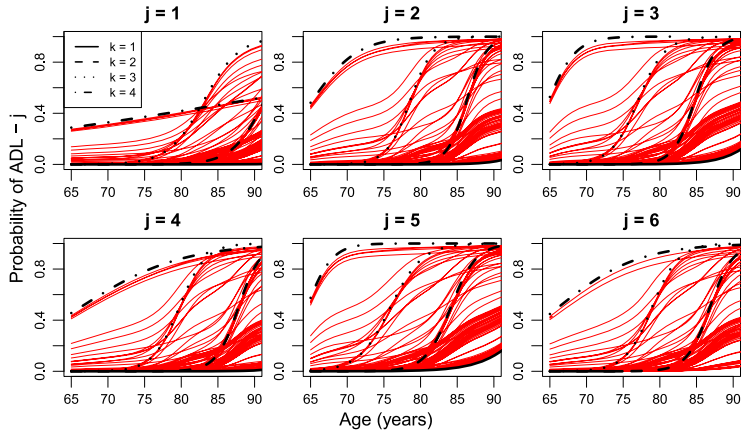}

\caption{Individual-level mixture of trajectories for model with $K=4$
extreme profiles for each ADL. Extreme trajectories are represented
with thick lines and a random sample of 100 individual posterior
trajectory curves are plotted using thin lines.}
\label{figbasicmusclek4}
\end{figure}

To better understand the way TGoM models handle individual-level
heterogeneity, it is instructive to visualize, in addition to the
extreme trajectories, the actual individual trajectories that result
from the individual-level mixing, $\lambda^{(i)}_{jk}(\tau)$. Plots
in Figures~\ref{figbasicmusclek3} and~\ref{figbasicmusclek4}
show a random sample of 100 such curves, overlaid over the three
extreme trajectory curves, for each ADL, under the model with $K=3$ and
$K=4$ extreme profiles, respectively. We see that most of the
individual curves cluster in the vicinity of the extreme curves. This
is expected, given the small value of the concentration parameter,
$\alpha_0$. However, we also see that a significant portion of the
individual curves lies somewhere in between extremes, exhibiting
trajectories that are the product of the interaction of more than one
extreme. In particular, we observe a fair number of individual
trajectories that fall in between extremes $k=1$ and $k=2$.
These trajectories form a somewhat homogeneous cluster different from
the extreme profiles. Nonetheless, the TGoM model has been able to
accommodate them as a combination of (mainly) profiles $k=1$ and $k=2$,
without needing to create a whole new category for them.
This behavior is what gives TGoM models the flexibility to accommodate
complex individual heterogeneity while at the same time producing
meaningful and interpretable summaries. Different from traditional LTMs
[\citet{nagin1999,connor2006}], which require that individuals
follow one and only one of the typical trajectories, this approach
allows them to depart from the main tendencies, but not too much, thus
retaining interpretability.

%f4 #&#

%
%f5 #&#

%s5.1.2 #&#
\subsubsection{Multivariate model diagnostics}
\label{secfit}
TGoM models explicitly model indi\-vidual-level dependency between
disability outcomes, both longitudinally and between ADLs, with the
help of an individual-level Mixed Membership structure.
In order to investigate empirically how TGoMs handle this dependency,
we evaluate posterior univariate and multivariate out-of-sample
predictive quantities. We define
%
%e5.2 #&#
%e5.3 #&#
%e5.4 #&#
%e5.5 #&#
\begin{eqnarray}
\label{eqpredictiongomfirst} \phi^{K}_{ijt} &=& \Pr\bigl(y^*_{ijt}
= y_{ijt} \mid \mathcal{D}, K\bigr),
\\
\phi^K_{ij} &=& \Pr\bigl(y^*_{ijt} =
y_{ijt}, \mbox{ for all }t\mid \mathcal{D}, K\bigr),
\\
\phi^K_{it} &=& \Pr\bigl(y^*_{ijt} =
y_{ijt}, \mbox{ for all }j\mid \mathcal{D}, K\bigr),
\\
\label{eqpredictiongomlast} \phi^K_{i} &=& \Pr\bigl(y^*_{ijt}
= y_{ijt}, \mbox{for all }t\mbox{ and all }j\mid \mathcal{D}, K
\bigr),
\end{eqnarray}
where $y^*_{ijt}$ is the posterior predictive outcome of individual $i$
in ADL-$j$ at wave~$t$, $\mathcal{D}$ are the NLTCS data,
and $K$ refers to the number of extreme profiles.
Thus, for individual $i$, $\phi^K_{ijt}$ is the (univariate) posterior
probability of correctly predicting outcome $y_{ijt}$ using a TGoM with
$K$ extreme profiles;\vspace*{2pt} $\phi^K_{ij}$ is the probability of \textit{simultaneously} correctly predicting the whole sequence of responses to
ADL-$j$, at all waves; $\phi^K_{jt}$ is the corresponding probability
of correctly predicting all the ADLs at wave $t$; and $\phi_i$ is the
probability of simultaneously correctly predicting all the responses of
an individual. In order to estimate the out-of-sample predictive
performance of our models, we compute all these quantities using a
4-fold cross-validation scheme [\citet
{hastie2009elements,airoldi2010pnas}].

As a comparison we also fit a model that assumes stochastic
independence between univariate outcomes, given age. We fit six (one
for each ADL) nonparametric logistic regressions of $\pi_{ijt} = \Pr
(y_{ijt}=1)$ on age, using Generalized Additive Models [GAM;
\citet{hastie2009elements}, Chapter~9]. We use this model as a
reference for assessing how our models handle the multivariate
structure present in the data. To this end,\vspace*{1pt} we compute quantities
analogous to \eqref{eqpredictiongomfirst}--\eqref{eqpredictiongomlast}:
$\phi^{\GAM}_{ijt} = \bernoullid(y_{ijt}\mid \hat\pi_{ijt})$, $\phi
^{\GAM}_{ij} = \prod_t \phi^{\GAM}_{ijt}$, $\phi^{\GAM}_{it} = \prod_j
\phi^{\GAM}_{ijt}$, and $\phi^{\GAM}_{i} = \prod_{j,t} \phi
^{\GAM}_{ijt}$, where $\hat{\pi}_{ijt}$ is the fitted value of $\pi
_{ijt}$. We compute these quantities using the same 4-fold
cross-validation scheme we use for the TGoM quantities.

Table~\ref{tabmultivariatepredictionoutsample} shows the 4-fold
cross-validated means (over all their subindexes) of $\phi_{ijt}$,
$\phi_{ij}$, $\phi_{it}$ and $\phi_{i}$, for TGoM with $K=1,2,\ldots,5$
extreme profiles and for the logistic GAM models. We take these numbers
as estimates of the corresponding file-level rates of correct
predictions for each model. We note that both TGoM and GAM models have
similar univariate prediction rates,
of around $80\%$, slightly favoring GAM. However, most of the joint
prediction rates of the TGoM models with $K>1$ are substantially better
than the alternative. In particular, TGoM correct prediction rates for
complete individual outcomes vectors, $\overline{\phi_i}$, are between
$41.4\%$ and $45.4\%$, while for the GAM model it drops down to $24.7\%
$. We also observe that multivariate prediction rates using TGoM models
tend to be much closer to their univariate prediction rates than the
corresponding quantities using the GAM alternative. For instance, the
ratio $\overline{\phi_{i}}/ \overline{\phi_{ijt}}$ (numbers between
parenthesis in the 5th column of Table~\ref
{tabmultivariatepredictionoutsample}) ranges from $51.6\%$ to $56.8\%
$ for TGoM models ($K=2$ and $K=5$, resp.), while for GAM it
falls down to $30.4\%$. We finally observe that estimates with the TGoM
model with $K=1$ are almost identical to those obtained with GAM. This
is because fitting TGoMs with only one extreme profile ($K=1$) is
equivalent to fitting $J$ independent logistic regressions of the
response variable (ADLs) on the predictors ($\Age$).

An interesting feature in Table~\ref
{tabmultivariatepredictionoutsample} is that longitudinal
predictions ($\overline{\phi_{ij}}$) are better than cross-sectional
predictions ($\overline{\phi_{it}}$) for all models. This
can be explained noting that
both the TGoM and GAM approaches exploit the extra longitudinal
information provided by the vectors of individuals' ages. By contrast,
when modeling the multivariate cross-sectional structure, TGoM relies
only on Mixed Membership and GAM only on independence given age.
Nonetheless, the comparison between the two modeling approaches still
favors TGoM models.

The conclusion of this prediction exercise is that TGoM models with
more than one extreme profile do capture a large portion of the
multivariate structure present in the data, both longitudinally and
cross-sectionally.

%
%t3 #&#
\begin{table}%[]
\tabcolsep=0pt
\caption{Out-of-sample rates of univariate and multivariate correct
predictions for TGoM and nonparametric logistic regression models.
Percentages between parentheses are the ratio of each entry with
respect to its corresponding univariate correct prediction rates, $\overline{\phi_{ijt}}$}\label{tabmultivariatepredictionoutsample}
\begin{tabular*}{\tablewidth}{@{\extracolsep{\fill}}@{}lcccc@{}}
\hline
\textbf{Model} &\multicolumn{1}{c}{$\bolds{\overline{\phi_{ijt}}}$}& \multicolumn{1}{c}{$\bolds{\overline{\phi_{ij}}}$} &
\multicolumn{1}{c}{$\bolds{\overline{\phi_{it}}}$} & \multicolumn{1}{c@{}}{$\bolds{\overline{\phi_{i}}}$}\\
\hline
TGoM $K=1$ & 0.811 (100\%) & 0.644 (79.4\%) & 0.452 (55.7\%) & 0.251 (30.9\%)\\
TGoM $K=2$ & 0.803 (100\%) & 0.666 (82.9\%) & 0.567 (70.6\%) & 0.414 (51.6\%)\\
TGoM $K=3$ & 0.802 (100\%) & 0.668 (83.3\%) & 0.593 (73.9\%) & 0.440 (54.9\%)\\
TGoM $K=4$ & 0.801 (100\%) & 0.668 (83.4\%) & 0.605 (75.5\%) & 0.451 (56.3\%)\\
TGoM $K=5$ & 0.799 (100\%) & 0.664 (83.1\%) & 0.607 (76.0\%) & 0.454 (56.8\%)\\
GAM-logistic & 0.812 (100\%) & 0.645 (79.4\%) & 0.451 (55.5\%) & 0.247 (30.4\%)\\
\hline
\end{tabular*}
\end{table}

%s5.2 #&#
\subsection{Fitting the cohort extensions}
We have fitted the model with extensions to handle cohort information
to the NLTCS data using the MCMC algorithms in Appendix~\ref
{secmcmcgrouped}, for $K=2,3$ and 4 extreme profiles.

The main objective of the analysis with this model is to compare the
underlying distribution of the membership vectors conditional on
generational groups, as a way of assessing differences in the ways of
aging between different cohorts. We do this by directly comparing the
parameters of these distributions for each generational group $\gamma
\in\Gamma$, $\balpha^\gamma$, and interpreting them with respect to
the common extreme trajectories, defined by the parameters
$(\bolds\beta_{jk})$.

\begin{figure}

\includegraphics{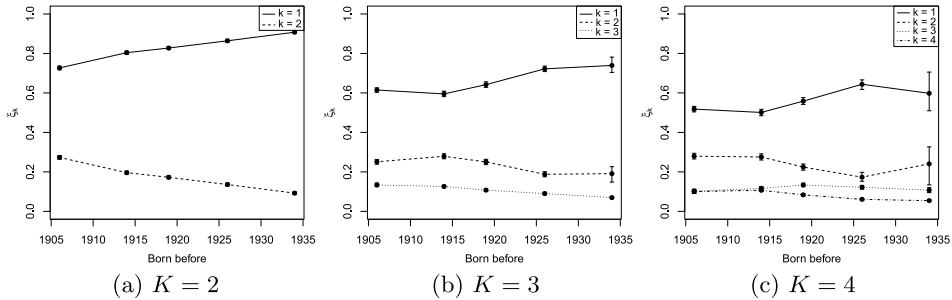}

\caption{Evolution of the parameter vector $\bxi$ across different
generations for models with $K = 2, 3$ and 4 extreme profiles. The
error bars show 95\% equal tail posterior credible intervals associated
with the kth component of the vector $\bxi$.}
\label{figgenerationalxis}
\end{figure}

Figure~\ref{figgenerationalxis} shows the estimates (posterior
means) of the components of the vector~$\bxi$ for models with $K=2,3$
and $4$ extreme profiles, for each cohort. For each generational group,
$\gamma_c$, the sequence of values of $\xi^{\gamma_c}_k$ are linked
with lines. Reading from left to right, these sequences indicate the
evolution of the relative weight of the $k$th component in each cohort
as we shift our attention from older to younger cohorts. Posterior
estimates of the common extreme profile parameters, $(\bolds\beta
_{jk})$, are very similar to those computed using the basic TGoM model
(see the supplementary material [\citet{supp}] for details), so we can safely refer to them when
discussing extreme profiles.

%
%f6 #&#

The most salient feature in Figure~\ref{figgenerationalxis} is the
increasing monotonicity of the relative importance of the first
component ($k=1$) in each cohort as we consider younger and younger
cohorts, that is, $\xi^{\gamma_1}_{1}<\xi^{\gamma_2}_{1}<\cdots <\xi
^{\gamma_5}_{1}$. This is especially clear in models with $K=2$ and
$K=3$. In the model with $K=4$, because of the high posterior
dispersion, it is not clear if the youngest generation actually follows
this pattern. A likely explanation for this uncertainty is the lack of
data for ages past $78$ years old in cohort 5.

This trend tells us that, as we consider newer cohorts, their members
tend to be increasingly close to profile $k=1$. This profile
corresponds to the healthiest aging progression, with extremely low
probability of acquiring disabilities until very advanced ages, as can
be observed in Figure~\ref{figbasicprofiles}. Thus, we conclude that
younger generations tend to have healthier ways of aging compared to
their elders.

%s6 #&#
\section{Discussion}\label{secdiscussion}
The methods we propose and apply here have several desirable features.
First, they produce meaningful and easy-to-interpret summaries of the
main temporal trends in the population. In this application, these
summaries---the extreme trajectories---isolate typical ways of
progressing into disability and allow a simplified analysis of the
longitudinal patterns. Second, they allow a simple, but not
oversimplified, characterization of the individual heterogeneity in
terms of the extreme trajectories. This keeps the extreme profile
characterizations simple, while still allowing the representation of
complex individual trajectories. Finally, the model's extensions allow
comparisons between groups of individuals defined by given static
characteristics. In this application it enables the separation of time-dependent effects that depend on age from those dependent on birth cohort.

The results obtained through the application of our methods to the
NLTCS highlight some interesting characteristics of the data and, in
general, of the aging process in the U.S. All the models considered
here showed that most individuals are close to the ``healthy aging''
profile ($k=1$), whose associated extreme trajectories (for the 6 ADLs)
describe a practically disability-free life until very late ages (90$+$).
Then we find that profiles with trajectories that specify earlier
onsets of disability exhibit progressively less importance in the
population. This means that most people could be expected to have a
relatively disability-free old age and that very bad aging processes
are not so common.

When considering the effect of the birth cohort---estimating
simultaneously population-wide extreme profiles together with
individual membership conditional on cohort---we find a similar
situation. However, different generations have a different membership
composition: the relative importance of the ``healthy aging'' profile
($k=1$) experiences a monotonic increase when moving from older to
younger generations, to the detriment of all the other profiles. Thus,
the answer to the question ``do younger generations acquire
disabilities differently than older ones?'' appears to be affirmative.
Furthermore, it is so in a positive sense: not only do younger
generations acquire disabilities differently, they also acquire them
later. These findings are consistent with previous evidence showing a
decline in disability obtained from purely cross-sectional analyses
[\citet{mantonstallard1997,mantonetal2006}], from wave to
wave latent class transition analysis [\citet
{whiteerosheva2013}], and from latent trajectory analysis
[\citet{connor2006}].

So far declines in disability have been analyzed mostly from wave to
wave, either from uncorrelated cross-sectional samples as changes in
prevalence [\citet
{mantonstallard1997,mantonetal2006,mantonlambwu2007}] or from
longitudinal analysis as transitions between states [\citet
{stallard2005,whiteerosheva2013}]. Our approach, in contrast, is
not rooted on survey waves, nor does it directly assess changes in
prevalence of disability. Instead, it characterizes whole individual
life trajectories. It therefore enables direct comparisons across
different ways of aging.

An important issue that we have addressed informally here is choosing
the number of extreme profiles, $K$.
In Section~\ref{secfit} we noted that in general the out-of-sample
multivariate fit measures improved with model complexity, increasing
$K$, although the improvement was different depending on which
multivariate dimension we chose to analyze.
We also have observed that the least important profiles in models with
$K>4$ do not reveal informative trajectories. Furthermore, we note that
our conclusions do not really depend on an exact number of extreme
profiles. Therefore, we evaluate that in this case we do not need to
select a ``best'' model; instead, we have opted for reporting results
from several models, with numbers of extreme profiles ranging from
$K=2$ to $K=4$. Model selection, however, can be an important issue in
other applications. Possible approaches include the use of indexes such
as AIC [\citet{akaike1973information}] or BIC [\citet
{schwarz1978estimating}]---or their more computationally convenient
counterparts DIC [\citet{spiegelhalter2002bayesian}], AICM or
BICM [\citet{raftery2007estimating}]---although in this case the
difficulty in computing the integrated likelihood could make this
approach impractical.
Another approach is to use Bayesian nonparametric specification that
favor sparse representations, such as Dirichlet Process mixtures.
\citet{bhattacharya2012simplex} have proposed what is essentially
a nonparametric Mixed Membership model for categorical data which could
be adapted for this purpose.

As for model limitations, this analysis attributes all variability in
the data to a combination of random fluctuation, age effects, cohort
effects and Mixed Membership. Thus, it neglects other potential
systematic effects, some of which might be important either to capture
previously unaccounted variability or simply for better understanding
the underlying processes. For instance, it is well known [see, e.g.,
\citet
{ferruccietal1996progressivevscatastrophicdisabilitylongitudinal,manton2008phreview}]
that men and women follow different aging and mortality processes. One
natural way of accounting for nontime-dependent categorical covariates,
like gender or race, is to introduce them in the same way we introduced
the DOB covariate: as conditioners on the prior distribution of
individual membership. If the cells on the contingency table generated
by the cross-classification according to the covariates are well
populated, we can directly use the TGoM extensions from Section~\ref
{secgenerational}. If this is not the case, the joint covariate vector
can be smoothed using more complex prior specifications, such as those
proposed in \citet{bertolet2008} for the Grade of Membership model.

Another related limitation of these models is that they do not account
for mortality. In essence, these models correspond to what \citet
{kurland2005conditionalonbeingalive} and \citet
{kurland2009longitudinaltruncationbydeasth} call an ``immortal
cohort.'' This is of particular importance in the present application
because patterns of disability are usually tied to patterns of
mortality [\citet
{ferruccietal1996progressivevscatastrophicdisabilitylongitudinal,connor2006,whiteerosheva2013}]:
progression into more severe disability goes together with an increased
probability of death. One way of integrating mortality into this
framework is to extend the definition of extreme profiles to
characterize not only patterns of disability acquisition, but also of
survival. Such a joint model could be the topic of a future article.

\begin{appendix}
%s8 #&#
\section{MCMC sampler for the TGoM model}\label{secmcmcbasic}
In this appendix we present a Gibbs sampling algorithm for Bayesian
estimation of the TGoM model, also described in \citet
{manrique2013}. Following the discussion at the end of Section~\ref{secbasicconstruction},
we construct an algorithm for obtaining samples from the posterior
distribution of parameters in the augmented data model in
equation~\eqref{eqbasicaugmentedabstract}, which after
marginalizing $\mathbf{z}$ is equivalent to the TGoM model. This
posterior distribution is
%
%e8.1 #&#
\begin{equation}
p(\balpha, \bolds{\beta}, \bZ, \bg\mid \bY, \mathbf{Age}) \propto p(\balpha,
\bg, \bolds{\beta}) \prod_{i=1}^N
f^{\AUG} (\bY_i, \bZ_i \mid
\mathbf{Age}_i, \bbeta, \bg_i ),
\end{equation}
which following the
detailed specification from Section~\ref
{secbasiccontinuoustrajectories} is equivalent to
%
%e8.2 #&#
%e8.3 #&#
%e8.4 #&#
\begin{eqnarray}
&& p(\balpha, \bolds{\beta}, \bZ, \bg\mid \bY, \mathbf{Age})\nonumber
\\
&&\qquad  \propto \gammad(
\alpha_0 \mid a_\alpha, b_\alpha) \times
\dirichletd (\bxi\mid \mathbf{1}_K )
\nonumber\\[8pt]\\[-28pt]\nonumber
&&\quad\qquad{} \times{\prod_{i = 1}^N {\dirichletd(
\bg_i\mid \balpha)} } \times\prod_{j=1}^J
\prod_{k=1}^KN\bigl(\beta_{0jk}
\mid \mu_0, \sigma_0^2\bigr) \times N\bigl(
\beta_{1jk}\mid \mu_1, \sigma_1^2
\bigr)
\\
&&\quad\qquad{}\times\prod
_{i = 1}^N {} \prod
_{j = 1}^J \prod
_{t=1}^T
{ {g_{iz_{ijt}} \frac{{\exp(y_{ijt} \beta
_{0jz_{ijt}} + y_{ijt} \beta_{1jz_{ijt}} \Age_{it} )}}{
{1 + \exp(\beta_{0jz_{ijt}} + \beta_{1jz_{ijt}} \Age_{it} )}}} },\nonumber
\end{eqnarray}
with $\alpha_0 = \sum\alpha_k$ and $\bxi= (\alpha_1/\alpha
_0,\ldots,\alpha_K/\alpha_0)$. Parameters $a_\alpha$ and $b_\alpha$
are shape and inverse scale parameters, respectively.

A Gibbs sampling algorithm for obtaining samples from the joint
posterior distribution of $(\balpha, \bolds{\beta}, \bZ, \bg
)$ can be constructed as follows:
\begin{longlist}
\item[1. \textit{Sampling from} $\bZ$:]
For every $i \in\{1,\ldots,N\}, j \in\{1,\ldots,J\}$ and $t \in \{
1,\break \ldots,T\}$, sample
$z_{ijt}\mid \cdots  \sim\discreted(\{1,\ldots,K\}$, $(p_1, p_2, \ldots, p_K))$, with
\[
p_k \propto g_{ik} \frac{\exp [y_{ijt}(\beta_{0jk} +\beta
_{1jk} \Age_{it} ) ]}{1 + \exp(\beta_{0jk} +\beta_{1jk}
\Age_{it} )}
\]
for all $k \in\{1,\dots, K\}$.

\item[2. \textit{Sampling from} $\bolds\beta_{jk}$:]
let $\Xi= \{(i,t)\dvtx  z_{ijt}=k\}$ and assume that $\mu_0 = \mu_1 = 0$.
The full joint conditional distribution of $(\beta_{0jk}, \beta
_{1jk})$ is
\begin{eqnarray*}
&& p ( {\beta_{0jk}, \beta_{1jk} \mid \cdots } )
\\
&&\qquad  \propto
\frac{
\exp [
- ({\beta_{1jk}^2}/{2\sigma_1^2} +{\beta
_{0jk}^2}/{2\sigma_0^2}  )
+ \beta_{0jk}\sum_\Xi y_{ijt}
+ \beta_{1jk}\sum_\Xi \Age_{it}y_{ijt}
 ]
}{
\prod_\Xi [ 1 + \exp ( \beta_{0jk} + \beta
_{0jk}\Age_{it}  )  ] }.
\end{eqnarray*}
To sample from this distribution, we use a random walk Metropolis step:
\begin{enumerate}[(a)]
\item[(a)] Sample proposal values
$\beta^*_{0jk}\sim N(\beta_{0jk},\sigma_{\beta0}^2)$ and $\beta
^*_{1jk}\sim N(\beta_{1jk},\sigma_{\beta1}^2)$, where $\sigma
_{\beta0}^2$ and $\sigma_{\beta1}^2$ are tuning parameters.\vspace*{1pt}

\item[(b)] With probability
\begin{eqnarray}\label{eqbasicsamplerbetarejectionlastline}
r_M
&=& \min \biggl\{1, \prod
_\Xi{
\biggl[ {\frac{{1 + \exp [
{{\beta_{0jk}} + {\beta_{0jk}}\Age_{it}}  ]}}{
{1 + \exp [ {\beta^*_{0jk} + \beta^*_{0jk}\Age_{it}}
]}}} \biggr]}\nonumber
\\
&&\hspace*{32pt}{}\times \exp \biggl[ - \frac{{\beta^{*2}_{0jk} - \beta_{0jk}^2}}{
{2\sigma_0^2}} + \bigl( {
\beta^*_{0jk} - {\beta_{0jk}}} \bigr) \sum_\Xi y_{ijt} \biggr]
\\
&&\hspace*{32pt}{}\times \exp \biggl[ - \frac{{\beta^{*2}_{1jk} - \beta_{1jk}^2}}{
{2\sigma_1^2}} + \bigl( {
\beta^*_{1jk} - {\beta_{1jk}}} \bigr) \sum_\Xi y_{ijt}\Age_{it} \biggr] \biggr\},\nonumber
\end{eqnarray}
make $(\beta_{0jk}, \beta_{1jk}) = (\beta^*_{0jk}, \beta^*_{1jk})$.
Otherwise keep the current value.
\end{enumerate}

\item[3. \textit{Sampling from} $\bg_i$:]
\begin{eqnarray*}
&&\bg_i\mid \cdots  \mathop{\sim}\limits^{\indep} \dirichletd \biggl(\alpha
_1 + \sum_{j,t} I(z_{ijt} = 1),
\ldots, \alpha_K + \sum_{j,t}
I(z_{ijt} = K) \biggr).
\end{eqnarray*}

\item[4. \textit{Sampling from} $\balpha$:]
The full conditional distribution of $\balpha$,
\begin{eqnarray}%\label{eqbasicsampleralpha}
\nonumber
&&p(\balpha\mid \cdots )  \propto \alpha_0^{a_\alpha-1}
e^{-\alpha_0 b_\alpha} \times \biggl[\frac{\Gamma ( \alpha_0  )}{\prod_{k=1}^K\Gamma(\alpha_k)} \biggr]^N \prod
_{k=1}^K \Biggl[ \prod
_{i=1}^Ng_{ik} \Biggr]^{\alpha_k},
\end{eqnarray}
does not have any recognizable form. We use a Metropolis--Hastings step
similar to \citet{manriquefienberg2008GoMCR}:
\begin{enumerate}[(a)]
\item[(a)] Obtain the proposal
\[
\balpha^* = \bigl(\alpha^*_1, \alpha^*_2,\ldots,
\alpha^*_K\bigr)\qquad\mbox{with }\alpha^*_k \mathop{\sim}\limits^{\indep} \operatorname{lognormal}\bigl(\log\alpha_k, \sigma^2\bigr).
\]

\item[(b)] Let $\alpha^*_0 = \sum_{k=1}^K\alpha^*_k$. With probability
\begin{eqnarray*}
r&=&\min \Biggl\{1,e^{ - a_\alpha(\alpha_0 ^* - \alpha_0 )} \biggl(
\frac{\alpha_0 ^*}{\alpha_0} \biggr)^{b_\alpha- 1} \Biggl( \prod
_{k = 1}^K
\frac{\alpha_k ^* }{\alpha_k } \Biggr)
\\
&&\hspace*{31pt}{}\times \Biggl[ \frac{\Gamma(\alpha_0 ^* )}{\Gamma(\alpha_0 )} \prod
_{k = 1}^K
\frac{\Gamma(\alpha_k )}{\Gamma(\alpha_k^*)} \Biggr]^N \prod
_{k = 1}^K
{ \Biggl( {\prod
_{i = 1}^N {g_{ik} } }
\Biggr)^{\alpha_k ^* - \alpha_k } } \Biggr\},
\end{eqnarray*}
make $\balpha= \balpha^*$. Otherwise keep the current value.
Obtain $(\alpha_0, \bxi)$ by making $\alpha_0=\sum_{k=1}^K \alpha
_k$ and $\xi_k = \alpha_k / \alpha_0$, for all $k=1,\ldots,K$.
\end{enumerate}
\end{longlist}

%s9 #&#
\section{Fitting the generational extension}\label{secmcmcgrouped}
The only difference between the posterior distributions of the basic
and the extended TGoM models is the distribution of $\bg_i\mid \balpha$.
Thus, we only have to adapt steps~3~and~4 in the previous algorithm by
replacing $\prod_{i=1}^N p(\bg_i\mid \balpha)$ with
%
%e9.1 #&#
\begin{equation}\label{eqbasicsampleralpha}
\prod_{i=1}^N p \bigl(\bg_i
\mid \balpha(\DOB_i) \bigr) = \prod_{\gamma
\in \Gamma}
\prod_{i=1}^N \bigl[ p \bigl(
\bg_i\mid \balpha^\gamma \bigr) \bigr]^{I(\DOB_i \in\gamma)}.
\end{equation}
Let $\gamma_i \in\Gamma$ be the unique interval from the partition
such that $\DOB_i \in\gamma_i$. We obtain an MCMC sampler for this
model by modifying steps 3 and 4 from the algorithm in Appendix~\ref{secmcmcbasic} with
the following:
\begin{longlist}[4$'$.]
\item[3$'$.] \textit{Sampling from} $\bg_i$:
\[
\bg_i\mid \cdots  \mathop{\sim}\limits
^{\indep} \dirichletd \biggl(
\alpha_1^{\gamma
_i}+\sum_{j,t}
I(z_{ijt}=1),\ldots,\alpha^{\gamma_i}_K+\sum
_{j,t} I(z_{ijt}=K) \biggr).
\]

\item[4$'$.] \textit{Sampling from} $\balpha$: let $\Xi_{\gamma}=\{i\dvtx
\gamma_i = \gamma\}$. The full conditional distribution of~$\balpha
^{\gamma}$ is
\[
p\bigl(\balpha^{\gamma}\mid \cdots \bigr) \propto \bigl(
\alpha_0^\gamma\bigr)^{a_\alpha-1} e^{-\alpha^\gamma_0
b_\alpha} \times
\biggl[\frac{\Gamma ( \alpha_0^\gamma )}{\prod_{k=1}^K\Gamma(\alpha_k^\gamma)} \biggr]^{\#(\Xi_\gamma)} \prod
_{k=1}^K \biggl[ \prod_{\Xi_\gamma}g_{ik}
\biggr]^{\alpha
_k^\gamma},
\]
where $\#(\Xi_\gamma)$ is the number of elements in the set $\Xi
_\gamma$.

This expression is similar to \eqref{eqbasicsampleralpha} in
Appendix~\ref{secmcmcbasic}. We thus adapt the procedure by
replacing $r$ in step~4 of the
algorithm with
\begin{eqnarray*}
r &=& \min \Biggl\{1,\exp \bigl[ - \tau\bigl(\alpha_{0} ^* -
\alpha_0^\gamma \bigr) \bigr] \Biggl( {\prod
_{k = 1}^K {\frac{{\alpha_k ^* }}{
{\alpha^\gamma_k} }} } \Biggr) \biggl( {
\frac{{\alpha_0 ^* }}{
{\alpha^\gamma_0 }}} \biggr)^{\tau- 1}
\\
&&\hspace*{31pt}{}\times \Biggl[ {\frac{{\Gamma(\alpha_0 ^* )}}{
{\Gamma(\alpha^\gamma_0 )}}\prod
_{k = 1}^K
{\frac{{\Gamma
(\alpha^\gamma_k )}}{
{\Gamma(\alpha_k^* )}}} } \Biggr]^{\#(\Xi_\gamma)} \prod
_{k = 1}^K { \biggl( {\prod
_{i \in\Xi_\gamma}
{g_{ik} } } \biggr)^{\alpha_k ^* - \alpha^\gamma_k } } \Biggr\}.
\end{eqnarray*}
\end{longlist}
\end{appendix}

%s7 #&#
\section*{Acknowledgments}
The author would like to thank Professor Steve Fienberg for his
guidance and encouragement while conducting this investigation,
Professor Jerry Reiter for his help and valuable suggestions during the
review process, and three anonymous referees and one Associate Editor
for their thorough critique and recommendations. This work is partially
based on material from the Ph.D. thesis of the author in the Department
of Statistics at Carnegie Mellon University. For a closely related
treatment of the model and data analysis see also \citet{manrique2013}.

\begin{supplement}[id=suppA]
\stitle{Supplement to ``Longitudinal Mixed Membership trajectory models for disability survey data''}
\slink[doi]{10.1214/14-AOAS769SUPP} %[doi,text={\cdots }] - jei reikia suskaldyti doi
\sdatatype{.pdf}
\sfilename{AOAS769\_supp.pdf}
\sdescription{Estimation using TGoM models with piecewise constant
trajectories and tables with posterior estimates for all the fitted models.}
\end{supplement}

% imsref loaded by linak, 2014-09-10 14:31:18
%
% imsref loaded by linak, 2014-09-11 09:25:42

\printaddresses
\end{document}